\documentclass[
reprint,
aps,
prd, 
superscriptaddress,
showpacs,
nofootinbib,
amsmath,
amssymb,
longbibliography 
]{revtex4-2}
\usepackage{slashed}
\usepackage{graphicx}
\usepackage{dcolumn}
\usepackage{bm}
\usepackage{hyperref}
\usepackage{color}
\usepackage{orcidlink}
\usepackage{filecontents}
\usepackage{float}
\usepackage{booktabs}
\usepackage{mathtools}
\usepackage{tikz}

\usetikzlibrary{arrows.meta, positioning, shapes.geometric, calc, fit, backgrounds}

\usepackage{xcolor}
\definecolor{darkblue}{rgb}{0.0, 0.0, 0.55}
\definecolor{darkred}{rgb}{0.55, 0.0, 0.0}

\hypersetup{
	colorlinks=true,
	linkcolor=darkblue, 
	citecolor=darkred,  
	urlcolor=darkblue   
}

\begin{document}
	
	\title{Twisting Small-$x$ Gluon Tomography with Orbital Angular Momentum}
	
	\author{Wei Kou\orcidlink{0000-0002-4152-2150}}
	\email{kouwei@impcas.ac.cn}
	\affiliation{Institute of Modern Physics, Chinese Academy of Sciences, Lanzhou 730000, Gansu Province, China}
	\affiliation{Southern Center for Nuclear Science Theory (SCNT), Institute of Modern Physics, Chinese Academy of Sciences, Huizhou 516000, Guangdong Province, China}
	\affiliation{School of Nuclear Science and Technology, University of Chinese Academy of Sciences, Beijing 100049, China}
	\affiliation{State Key Laboratory of Heavy Ion Science and Technology, Institute of Modern Physics, Chinese Academy of Sciences, Lanzhou 730000, Gansu Province, China}
	
	\author{Xurong Chen}
	\email{xchen@impcas.ac.cn}
	\affiliation{Institute of Modern Physics, Chinese Academy of Sciences, Lanzhou 730000, Gansu Province, China}
	\affiliation{Southern Center for Nuclear Science Theory (SCNT), Institute of Modern Physics, Chinese Academy of Sciences, Huizhou 516000, Guangdong Province, China}
	\affiliation{School of Nuclear Science and Technology, University of Chinese Academy of Sciences, Beijing 100049, China}
	\affiliation{State Key Laboratory of Heavy Ion Science and Technology, Institute of Modern Physics, Chinese Academy of Sciences, Lanzhou 730000, Gansu Province, China}
	
	
	
	\begin{abstract}
		We propose an orbital-angular-momentum-resolved extension of small-$x$ gluon tomography in hard diffractive dijet deep inelastic scattering.  In the standard plane-wave setup, the elliptic correlation between the dijet relative momentum and the target recoil probes the elliptic component of the small-$x$ gluon Wigner distribution through a fixed transverse readout.  We show that replacing the plane-wave lepton current by a twisted wave-packet current promotes this readout into a tunable OAM--Bessel projection kernel.  The exchanged virtual photon is not treated as an asymptotic vortex particle; the OAM dependence enters through the transverse structure of the lepton electromagnetic current.  The resulting observable is a family of $M$-resolved elliptic correlations $A_2^{(M)}(q_T,R_\gamma)$, where $q_T$ denotes the transverse Bessel scale of the projection kernel, not the transverse momentum of the exchanged photon.  We derive the normalized linear response and show that it contains a subtraction from the deformation of the total diffractive rate.  Consequently, the elliptic response can vanish while the diffractive rate remains finite.  This finite-rate null is a normalized projection zero of the response to the target elliptic gluon component, not a disappearance of diffraction.  It is not available as a tunable mode zero in the standard single plane-wave readout, and provides an external projection basis in which the response to the same small-$x$ elliptic geometry can be probed with either sign or tuned to zero.
		
	\end{abstract}

	\maketitle

	\section{Introduction}
	\label{sec:introduction}
	
	A central goal of hadron physics is to image the quark and gluon structure of matter beyond one-dimensional parton densities.  Quantum phase-space distributions, such as Wigner distributions and generalized transverse-momentum-dependent distributions, provide a unified description of partonic motion in transverse momentum and transverse position, and are closely tied to orbital angular momentum in quantum chromodynamics (QCD)~\cite{Ji:2003ak,Belitsky:2003nz,Lorce:2011kd}.  In the gluon sector, related phase-space observables have been proposed as a route to gluon orbital angular momentum at the Electron-Ion Collider (EIC) and at small $x$~\cite{Ji:2016jgn,Hatta:2016aoc}.  The EIC will make such multidimensional imaging especially relevant in the gluon-dominated regime, where nonlinear QCD dynamics becomes experimentally accessible~\cite{Accardi:2012qut,AbdulKhalek:2021gbh}.  A key step was the proposal that correlated hard diffractive dijet production in deep inelastic scattering (DIS) can probe the elliptic component of the small-$x$ gluon Wigner distribution through the azimuthal correlation between the dijet relative momentum and the target recoil~\cite{Hatta:2016dxp}.  Complementary small-$x$ gluon tomography has also been discussed in deeply virtual Compton scattering~\cite{Hatta:2017cte}.
	
	Diffractive dijet tomography has since been developed in the color-glass-condensate and dipole formalisms.  In these descriptions, the virtual photon fluctuates into a compact quark-antiquark dipole whose coherent scattering from the target encodes both transverse-size and impact-parameter information.  The resulting $\cos2(\phi_P-\phi_\Delta)$ modulation is therefore not merely a final-state jet correlation, but a projection of the elliptic angular structure of the small-$x$ gluon field~\cite{Zhou:2016rnt,Hagiwara:2017fye,Altinoluk:2015dpi,Mantysaari:2019csc,Salazar:2019ncp}.  Saturation effects, multigluon correlations, soft-gluon radiation, and realistic EIC observables have further established diffractive dijets as a controlled channel for small-$x$ gluon imaging~\cite{Mantysaari:2019hkq,Hatta:2019ixj,Boer:2021upt,Iancu:2021rup,Hatta:2022lzj,Hatta:2024ocp,Shao:2024gri}.  In the standard formulation, however, the transverse readout is fixed by the plane-wave lepton kinematics.
	
	A separate development has shown that scattering can acquire qualitatively new information when the incoming or outgoing particles are prepared as wave packets carrying orbital angular momentum (OAM).  Vortex photons and electrons possess a transverse phase winding and a ring-like momentum distribution, allowing scattering amplitudes to sample phase, impact-parameter, and angular structures that are averaged out in the plane-wave limit~\cite{Jentschura:2010ap,Ivanov:2011bv,Ivanov:2011aa,Ivanov:2016oue,Bliokh:2017uvr,Lloyd:2017dob}.  Although most established vortex-electron realizations are at comparatively low energies, recent theoretical progress has clarified how vortex-state methods may be extended toward high-energy collisions, including diagnostics, threshold effects, and possible routes to ultrarelativistic vortex leptons~\cite{Ivanov:2022jzh,Karlovets:2022jym,Liu:2022nei,Li:2024fek,Ababekri:2024lgx}.  These developments suggest that OAM can serve not only as a beam property, but also as a controllable analyzer of spatial and angular correlations in scattering.
	
	In this Letter we combine these two ideas.  We ask whether the fixed plane-wave projection in small-$x$ diffractive dijet tomography can be promoted to an OAM-resolved projection by replacing the plane-wave lepton current with a twisted wave-packet current.  The exchanged virtual photon is not an asymptotic particle and cannot be prepared as a free vortex photon.  The phrase ``twisted virtual photon'' is used only as a shorthand for an OAM-projected electromagnetic exchange kernel generated by the lepton current.  Thus the target gluon distribution is not modified; the lepton side supplies a new transverse phase and radial weight with which the same elliptic small-$x$ structure is read out.
	
	We show that this structured exchange turns the usual elliptic dijet correlation into a family of $M$-resolved observables.  Different OAM channels sample the same target elliptic component with different Bessel radial weights and azimuthal phases, producing responses with different magnitudes and even opposite signs.  The key result is a finite-rate null: after the proper rate normalization is included, the linear elliptic response can vanish while the baseline diffractive rate remains nonzero.  This null is not a zero of the diffractive process or of the target dipole amplitude.  It is a normalized projection zero of the response to the elliptic gluon component, not a zero of the component itself, and it is not present as a tunable mode zero in the standard single plane-wave readout.

	\section{OAM-projected diffractive dijet DIS}
	\label{sec:oam-projected-dis}
	
\begin{figure}[htbp]
		\centering
		\includegraphics[width=\linewidth]{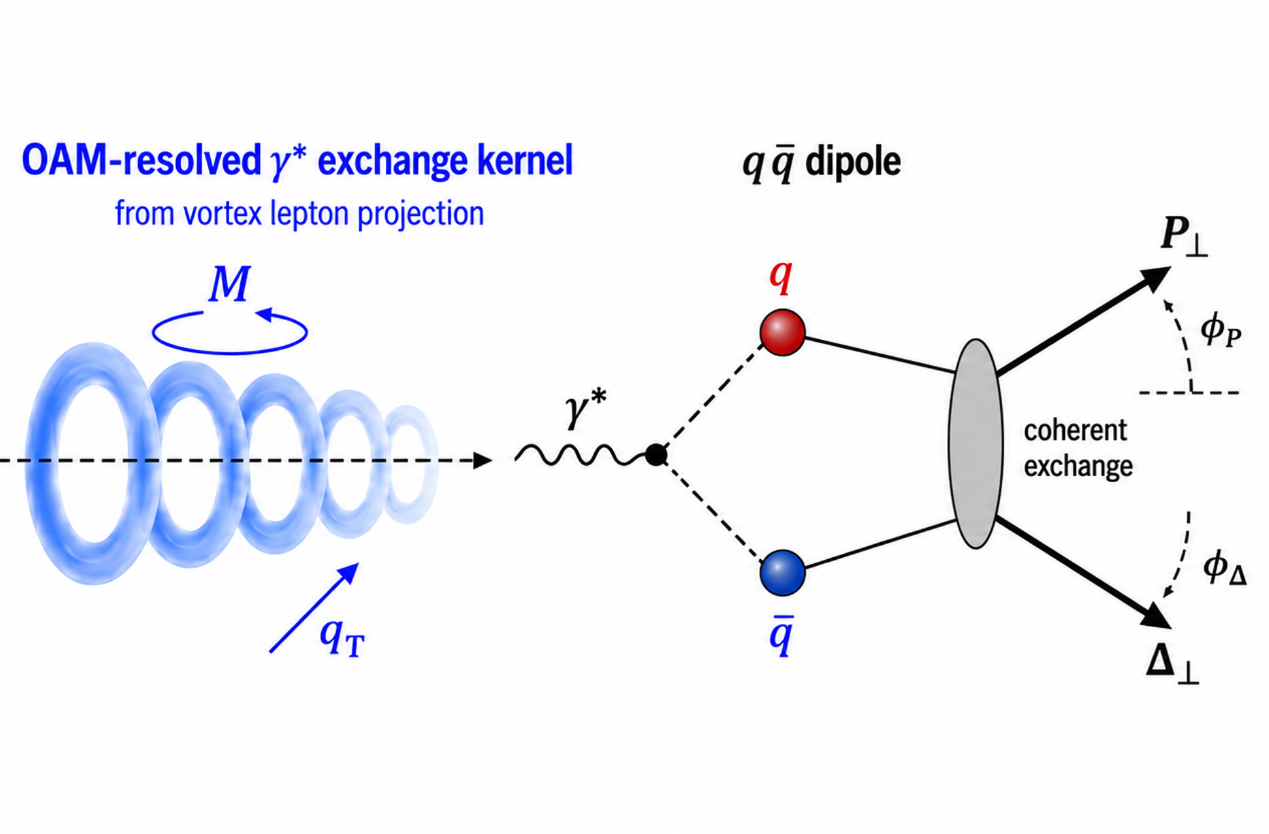}
		\caption{
			OAM-projected diffractive dijet DIS at small $x$.
			A vortex lepton projection induces an $M$-resolved electromagnetic exchange kernel, converting the usual plane-wave readout of the elliptic gluon distribution into a tunable transverse projection.
			Here $q_T$ labels the transverse Bessel scale of the projected kernel, while the DIS hard scale remains fixed by the photon virtuality $Q^2$.
			The measured azimuthal correlation is defined by the dijet relative momentum $\bm P_\perp$ and the recoil momentum $\bm\Delta_\perp$.
		}
		\label{fig:schematic}
	\end{figure}

	Diffractive dijet production in DIS provides a controlled channel for accessing the transverse geometry of small-$x$ gluons: the virtual photon creates a $q\bar q$ dipole, the dipole scatters coherently from the target color field, and the azimuthal correlation between the dijet momentum $\bm P_\perp$ and the recoil momentum $\bm\Delta_\perp$ projects the elliptic gluon structure.  In the plane-wave formulation this transverse projection is fixed by the external kinematics.  Here we introduce an externally controlled transverse projection by projecting the lepton current onto a vortex wave packet carrying OAM $M$, as sketched in Fig.~\ref{fig:schematic}. The resulting ``twisted virtual photon'' is not an asymptotic particle, but an $M$-resolved electromagnetic exchange kernel generated by the lepton current; the target distribution and the central DIS variables are kept unchanged.  A component-wise gauge-invariant construction of this projected current is given in Sec.~\ref{sec:sm-current} of the Supplemental Material.

	We work at fixed central DIS kinematics.  The hard scale and the small-$x$ evolution variable are specified by a central photon virtuality $Q_0^2$ and a target momentum fraction $x_{g0}$, which enter the photon splitting kernel and the dipole amplitude, respectively.  The OAM projection does not redefine either variable.  It changes only the transverse weight with which the same small-$x$ target field is sampled.  For a finite-width vortex wave packet, individual plane-wave components carry slightly different momenta and hence slightly different values of $Q^2$ and $x_g$~\cite{Ivanov:2011bv,Ivanov:2011aa,Karlovets:2016jrd,Ivanov:2022jzh}. In the narrow-packet regime considered here, these kinematic-smearing effects are treated as width-suppressed corrections around $(x_{g0},Q_0^2)$ and are separated from the OAM--Bessel projection.  The corresponding fixed-central-kinematics construction is summarized in Sec.~\ref{sec:sm-current} of the Supplemental Material.

	The transverse projection is implemented by replacing the plane-wave transverse phase with a Bessel-type wave-packet weight~\cite{Ivanov:2011bv,Ivanov:2011aa}.  We use $M$ to denote the OAM channel and $q_T$ to denote the transverse Bessel scale carried by the projected exchange kernel.  This is a mode parameter rather than an event-level transverse momentum: $q_T$ is an inverse transverse length that fixes the radial nodes of the OAM--Bessel weight, and should not be confused with the photon virtuality, the photon transverse momentum, or the recoil momentum.  The DIS hard scale is still $Q^2=-q^2$, while the target recoil is denoted by $\bm\Delta_\perp$.  At the level of the transverse projection used here, the kernel contains an azimuthal phase $e^{iM\phi_R}$, a radial Bessel weight $J_M(q_T R)$, and a finite transverse envelope.  Different values of $M$ and $q_T$ therefore probe the same target at the same central $(x_{g0},Q_0^2)$, but with different radial nodes and angular phases.

	On the target side we keep the standard small-$x$ dipole description, rooted in the high-density color-field picture and its small-$x$ evolution~\cite{McLerran:1993ni,McLerran:1993ka,JalilianMarian:1997jx,Balitsky:1995ub,Kovchegov:1999yj}.  The virtual photon splitting fixes the short-distance $q\bar q$ configuration, while the coherent interaction with the target is encoded in the dipole amplitude at rapidity $Y_0=\ln(1/x_{g0})$.  The relation between dipole scattering and small-$x$ gluon distributions provides the basis for using diffractive dijets as a gluon-imaging observable~\cite{Dominguez:2011wm,Hatta:2016dxp,Zhou:2016rnt,Altinoluk:2015dpi,Mantysaari:2019csc,Salazar:2019ncp}.  The role of the OAM projection is not to define a new target distribution, but to change the transverse overlap between the lepton-induced exchange kernel and the same dipole scattering amplitude.

	This construction turns the usual elliptic dijet modulation into an $M$-resolved observable.  For each OAM channel, the measured final state is unchanged, but the lepton current imposes a different transverse projection.  The plane-wave result is recovered when the OAM phase is removed and the transverse profile becomes broad, corresponding to $M=0$, $q_T\to0$, and $R_\gamma\to\infty$.  Away from this limit, the Bessel nodes and the azimuthal phase can increase, reduce, or change the sign of the response to the elliptic dipole component.  We now define this response quantitatively.

	\section{OAM-resolved elliptic response}
	\label{sec:oam-response}
	
	In plane-wave diffractive dijet tomography, the elliptic gluon structure is accessed through the second harmonic of the angle between the dijet relative momentum and the recoil momentum, conventionally written as a $\cos2(\phi_P-\phi_\Delta)$ modulation~\cite{Hatta:2016dxp,Zhou:2016rnt,Mantysaari:2019csc}.  With the OAM projection imposed on the lepton current, the same diffractive final state gives an $M$-dependent angular distribution.  We define the corresponding second-harmonic moment as
	\[
	A_2^{(M)}
	=
	\frac{
		\int d\phi_P\,
		\cos 2(\phi_P-\phi_\Delta)\,
		d\sigma_M/d\phi_P
	}{
		\int d\phi_P\,
		d\sigma_M/d\phi_P
	},
	\]
	where $d\sigma_M$ is the diffractive dijet cross section evaluated with the OAM-projected exchange kernel.
	
	To isolate the response to the target elliptic structure, we expand the OAM-resolved angular distribution in the elliptic deformation parameter $\epsilon_2$,
	\[
	\frac{d\sigma_M}{d\phi_P}
	=
	\frac{d\sigma_{0,M}}{d\phi_P}
	+
	\epsilon_2
	\frac{d\sigma_{1,M}}{d\phi_P}
	+
	{\cal O}(\epsilon_2^2).
	\]
	Let $W_{i,M}$ and $\Sigma_{i,M}$ denote the corresponding second-harmonic numerator and total-rate moments, respectively.  The normalized correlation then has the expansion
	\[
	A_2^{(M)}
	=
	v_{2,0}^{(M)}
	+
	\epsilon_2 S_M
	+
	{\cal O}(\epsilon_2^2),
	\qquad
	v_{2,0}^{(M)}=\frac{W_{0,M}}{\Sigma_{0,M}},
	\]
	with
	\[
	S_M
	=
	\frac{W_{1,M}}{\Sigma_{0,M}}
	-
	v_{2,0}^{(M)}
	\frac{\Sigma_{1,M}}{\Sigma_{0,M}} .
	\]
	The second term is fixed by the normalization of the angular correlation: the elliptic deformation changes not only the weighted numerator, but also the total diffractive rate. The derivation of this normalized response is given in Sec.~\ref{sec:sm-response} of the Supplemental Material.
	
	The benchmark dipole input used for the numerical results is specified in Sec.~\ref{sec:sm-dipole} of the Supplemental Material. Figure~\ref{fig:oam-response} shows the response as a function of the transverse Bessel scale $q_T$.  The OAM channels do not simply rescale the plane-wave result.  Changing $M$ modifies the radial overlap between the exchange kernel and the elliptic dipole component, leading to different magnitudes and signs of $S_M(q_T)$.  The lower panel shows the rate-weighted diagnostic quantity $|S_M|\sqrt{\Sigma_0}$, which indicates where a sizable response is accompanied by a non-negligible baseline rate.  This quantity is not an independent normalized observable.  The figure therefore demonstrates that the lepton-side OAM mode changes the projected elliptic response, not only its overall normalization.
	
	\begin{figure}[htbp]
		\centering
		\includegraphics[width=\linewidth]{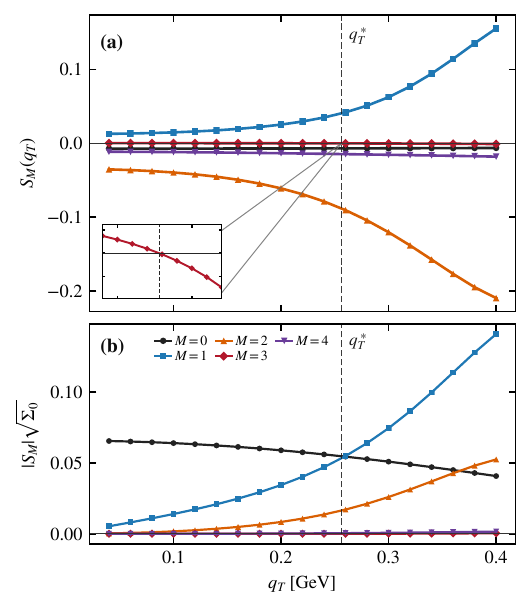}
		\caption{
			OAM-resolved elliptic response.
			Upper panel: linear response $S_M(q_T)$ for several OAM channels.
			Lower panel: rate-weighted diagnostic quantity $|S_M|\sqrt{\Sigma_0}$, shown to indicate where a sizable response is accompanied by a non-negligible baseline rate.
			The dashed line marks the finite-rate zero of the $M=3$ response.
		}
		\label{fig:oam-response}
	\end{figure}
	
\begin{figure}[htbp]
		\centering
		\includegraphics[width=\linewidth]{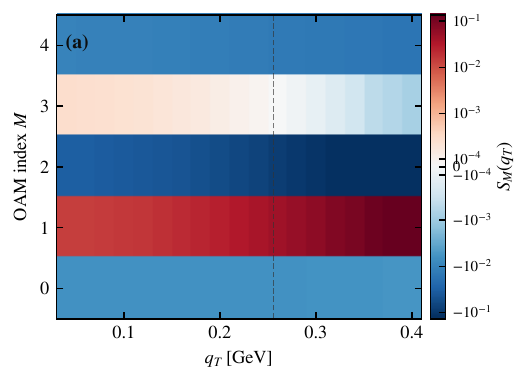}
		\caption{
			Response landscape in the $(M,q_T)$ plane.
			The color represents the linear elliptic response $S_M(q_T)$, shown on a symmetric logarithmic scale.
			The dashed line marks the finite-rate zero identified in the $M=3$ channel.
		}
		\label{fig:response-map}
	\end{figure}

	Figure~\ref{fig:response-map} summarizes the channel dependence in the $(M,q_T)$ plane.  The sign-changing pattern shows that the OAM projection does not merely select stronger or weaker copies of the same observable.  The radial nodes of the Bessel kernel reorganize the overlap with the elliptic dipole component as $M$ and $q_T$ are varied.  As a result, the linear response to the same target elliptic component can change sign or become strongly suppressed.  These suppressed regions are useful because they separate cancellation in the projected elliptic response from suppression of the overall diffractive rate.

\begin{figure}[htbp]
		\centering
		\includegraphics[width=\linewidth]{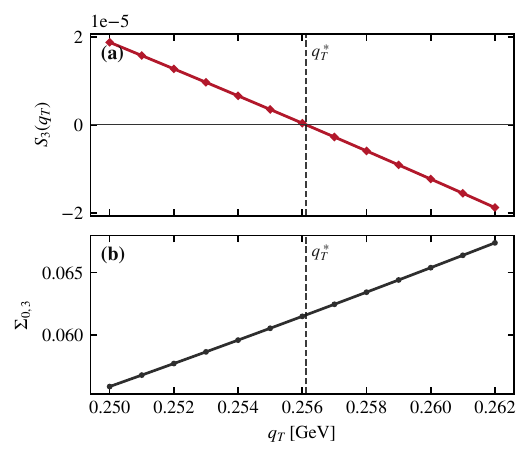}
		\caption{
			Finite-rate null of the OAM-resolved elliptic response.
			Upper panel: the $M=3$ response $S_3(q_T)$ changes sign and vanishes at $q_T^\ast\simeq0.256~{\rm GeV}$.
			Lower panel: the corresponding baseline rate $\Sigma_{0,3}$ remains finite at the same point.
		}
		\label{fig:finite-rate-null}
	\end{figure}

	The clearest consequence of the normalization term in $S_M$ is shown in Fig.~\ref{fig:finite-rate-null}.  For the $M=3$ channel, the linear elliptic response crosses zero at a finite Bessel scale while the baseline rate remains nonzero.  In the benchmark shown here, the zero occurs near
	\[
	q_T^\ast \simeq 0.256~{\rm GeV},
	\qquad
	\Sigma_{0,3}(q_T^\ast)\simeq 0.062 ,
	\]
	with $\Sigma_{0,3}$ expressed in the normalization used for the plotted rates.  This finite-rate null means that the diffractive process is still present, but its first-order sensitivity to the target elliptic component is canceled in the normalized OAM-projected response.  The cancellation would be missed if one identified the elliptic response only with the numerator correction $W_{1,M}$.  It is therefore a normalized observable effect controlled by the competition between the elliptic numerator and the deformation of the total diffractive rate.
	
	The linear-response interpretation is supported by a direct variation of the target elliptic deformation.  For representative values of $q_T$ on the two sides of the $M=3$ null, the full elliptic correlation follows the first-order prediction
	$A_2^{\rm full}-v_{2,0}^{(M)}=\epsilon_2 S_M$
	within the deformation range used here. This check, shown in Sec.~\ref{sec:sm-linearity} of the Supplemental Material, confirms that the zero in Fig.~\ref{fig:finite-rate-null} is not a finite-difference artifact, but a property of the normalized OAM-projected response.  Higher-order corrections in $\epsilon_2$ or kinematic smearing from a finite lepton wave packet can shift the detailed zero position, but do not affect the definition of the leading linear-response mechanism.The stability of this finite-rate response zero under representative variations of the benchmark input and of the OAM wave-packet width is summarized in Sec.~\ref{sec:sm-robustness} of the Supplemental Material.

	\section{Implications and outlook}
	\label{sec:outlook}
	
	The result above shows that OAM changes the projection of a fixed small-$x$ target matrix element, rather than defining a new target distribution.  The same elliptic dipole structure that appears in plane-wave diffractive dijet tomography is sampled with a different transverse phase and radial weight supplied by the lepton current.  The $M$-resolved observables therefore form different projections of the same gluon distribution.  The finite-rate null is the clearest example: the response to the elliptic component can be tuned to zero while the baseline diffractive rate remains nonzero.  The position of such a zero is model and wave-packet dependent, but the underlying mechanism is the normalized OAM--Bessel projection of the elliptic gluon component.
	
	A possible experimental implementation would require either control of the transverse mode of the incoming lepton or an OAM-sensitive projection of the outgoing lepton.  In both cases the relevant object is not a free vortex virtual photon, but an OAM-resolved lepton tensor whose transverse structure is transferred to the electromagnetic exchange.  High-energy vortex leptons remain experimentally challenging, although recent work has clarified possible routes for vortex-state generation, diagnostics, and high-energy wave-packet scattering~\cite{Ivanov:2022jzh,Karlovets:2022jym,Liu:2022nei,Li:2024fek,Ababekri:2024lgx}.  The present study should therefore be regarded as a proof-of-principle construction of an observable, rather than a complete feasibility assessment for a specific EIC implementation~\cite{Accardi:2012qut,AbdulKhalek:2021gbh}.
	
	A full phenomenological treatment should combine the OAM-projected lepton tensor with realistic jet reconstruction, soft-gluon radiation, and nuclear small-$x$ evolution, following the standards developed for plane-wave diffractive dijet observables~\cite{Hatta:2019ixj,Boer:2021upt,Iancu:2021rup,Hatta:2022lzj,Hatta:2024ocp,Shao:2024gri}.  Finite wave-packet effects can also be included by convoluting the plane-wave amplitude over the momentum distribution of the twisted lepton, which generates controlled smearing in $x_g$ and $Q^2$ around the central DIS kinematics.  These effects can shift the quantitative location of a response zero and modify the projected sensitivity, but they do not alter the definition of the leading OAM-resolved projection.  OAM-resolved diffractive dijets therefore provide a controlled way to test how the same small-$x$ elliptic gluon geometry is projected with different transverse lepton modes.

	\begin{acknowledgments}
		This work has been supported by the National Natural Science Foundation of China (Grant No.
		12547118), the Research Program of State Key Laboratory of Heavy Ion Science
		and Technology, Institute of Modern Physics, Chinese Academy of Sciences
		(Grant No. HIST2025CS08), and the National Key R\&D Program of China (Grant
		Nos. 2024YFE0109800 and 2024YFE0109802).
	\end{acknowledgments}
	
	\bibliographystyle{apsrev4-2}
	\bibliography{refs}
	
	\clearpage
	\onecolumngrid
	
	\setcounter{section}{0}
	\setcounter{equation}{0}
	\setcounter{figure}{0}
	\renewcommand{\thesection}{S\Roman{section}}
	\renewcommand{\theequation}{S\arabic{equation}}
	\renewcommand{\thefigure}{S\arabic{figure}}
	
	\begin{center}
		{\large \textbf{Supplemental Material}}\\[0.5em]
		{\large \textbf{Twisting Small-$x$ Gluon Tomography with Orbital Angular Momentum}}
	\end{center}
	
	\section{OAM-projected lepton current and fixed kinematics}
	\label{sec:sm-current}
	
	This section gives the minimal definitions used in the main text for the orbital-angular-momentum (OAM) projected electromagnetic exchange.  The exchanged virtual photon is not an asymptotic particle and is not assumed to be prepared as a free vortex photon.  The OAM dependence is introduced through the external lepton state, or equivalently through an OAM-projected lepton current.
	
	For each plane-wave component of the lepton wave packet, the electromagnetic current satisfies the Ward identity
	\begin{equation}
		q_\mu(k)\,J^\mu(k,k')=0,
		\qquad
		q(k)=k-k' .
		\label{eq:sm-current-conservation}
	\end{equation}
	The OAM-projected current is constructed as a coherent superposition of such components,
	\begin{equation}
		\begin{aligned}
			J_{M}^{\mu}
			=
			\int \frac{d^{2}k_\perp}{(2\pi)^2}\,
			a_M(k_\perp)\,
			e^{iM\phi_k}\,
			J^\mu(k,k') ,
		\end{aligned}
		\label{eq:sm-oam-current}
	\end{equation}
	where $M$ is the OAM index and $a_M(k_\perp)$ specifies the transverse wave-packet profile~\cite{Ivanov:2011bv,Ivanov:2011aa,Karlovets:2016jrd,Ivanov:2022jzh}.  At the exact wave-packet level, the Ward identity is imposed component by component with the corresponding momentum transfer $q(k)$.  In the fixed-central-kinematics approximation used in the main text, variations of $q(k)$ around the central momentum are treated consistently as width-suppressed corrections.
	
	The corresponding OAM-resolved amplitude can be written as
	\begin{equation}
		\begin{aligned}
			{\cal M}_M
			=
			\int \frac{d^2k_\perp}{(2\pi)^2}\,
			a_M(k_\perp)e^{iM\phi_k}\,
			{\cal M}_{\rm PW}
			\bigl[x_g(k),Q^2(k)\bigr] .
		\end{aligned}
		\label{eq:sm-wavepacket}
	\end{equation}
	Here $Q^2(k)=-q^2(k)$ and $x_g(k)$ are the hard scale and target momentum fraction associated with each plane-wave component.  In the narrow-packet regime, these variables are expanded around central values,
	\begin{equation}
		Q^2(k)=Q_0^2+\delta Q^2(k),
		\qquad
		Y(k)=Y_0+\delta Y(k),
		\qquad
		Y_0=\ln\frac{1}{x_{g0}} .
		\label{eq:sm-central-expansion}
	\end{equation}
	The main text retains the leading term at fixed $(x_{g0},Q_0^2)$.  The omitted terms describe kinematic smearing from the finite width of the lepton wave packet and are distinct from the OAM--Bessel projection.
	
	At the level of the transverse projection used in the main text, the exchange kernel is represented by
	\begin{equation}
		{\cal V}_{M}(\bm R)
		=
		{\cal N}_{M}\,
		e^{iM\phi_R}
		J_M(q_T R)
		\exp\left[
		-\frac{R^2}{2R_\gamma^2}
		\right] .
		\label{eq:sm-oam-kernel}
	\end{equation}
	The parameter $q_T$ is the transverse Bessel scale of the projected mode.  It is an inverse transverse length and should not be identified with the photon transverse momentum, the recoil momentum, or the photon virtuality.  The plane-wave limit is recovered by removing the OAM phase and taking the transverse profile to be infinitely broad,
	\begin{equation}
		M=0,
		\qquad
		q_T\to0,
		\qquad
		R_\gamma\to\infty .
		\label{eq:sm-plane-wave-limit}
	\end{equation}

	\section{Benchmark dipole input}
	\label{sec:sm-dipole}
	
	The numerical results in the main text use the same target input for all OAM channels.  The isotropic dipole amplitude is modeled by an impact-parameter-dependent generalized Golec-Biernat--W{\"u}sthoff (GBW)-type ansatz,
	\begin{subequations}
		\begin{align}
			Q_s^2(b,Y)
			&=
			Q_{s0}^2
			e^{\lambda_s Y}
			e^{-b^2/B_p},
			\label{eq:sm-Qs}
			\\
			N_0(r,b,Y)
			&=
			1-
			\exp\left[
			-\frac14
			\left(
			r^2Q_s^2(b,Y)
			\right)^\gamma
			\right] .
			\label{eq:sm-N0}
		\end{align}
	\end{subequations}
	This benchmark input is not used as a global impact-parameter saturation (IP-Sat) or impact-parameter dependent Color Glass Condensate (b-CGC) fit.  Its purpose is to provide a controlled target background on which the effect of the external OAM--Bessel projection can be isolated~\cite{GolecBiernat:1998js,Hatta:2016dxp,Zhou:2016rnt,Mantysaari:2019csc,Salazar:2019ncp}.
	
	The elliptic deformation is introduced as
	\begin{equation}
		\begin{aligned}
			N_Y(\bm r,\bm b)
			=
			N_0(r,b,Y)
			+
			\epsilon_2
			N_\epsilon(r,b,Y)
			\cos2(\phi_r-\phi_b)
			+
			{\cal O}(\epsilon_2^2).
		\end{aligned}
		\label{eq:sm-elliptic-dipole}
	\end{equation}
	The deformation profile is
	\begin{subequations}
		\begin{align}
			N_\epsilon(r,b,Y)
			&=
			2F_2(r,b)
			N_0(r,b,Y)
			\left[1-N_0(r,b,Y)\right],
			\label{eq:sm-Neps}
			\\
			F_2(r,b)
			&=
			\frac{r^2b^2}
			{(r^2+b^2+B_0^2)^2}.
			\label{eq:sm-F2}
		\end{align}
	\end{subequations}
	The factor $F_2$ suppresses the elliptic component when either the dipole size or the impact parameter vanishes, while $N_0(1-N_0)$ localizes the deformation near the transition between the dilute and saturated regimes.  Varying $M$, $q_T$, or $R_\gamma$ changes only the external projection kernel in Eq.~\eqref{eq:sm-oam-kernel}; the target functions $N_0$ and $N_\epsilon$ are kept fixed.

	\section{Normalized linear response}
	\label{sec:sm-response}
	
	For a fixed OAM channel, the angular distribution is expanded in the elliptic deformation parameter as
	\begin{equation}
		\frac{d\sigma_M}{d\phi_P}
		=
		\frac{d\sigma_{0,M}}{d\phi_P}
		+
		\epsilon_2
		\frac{d\sigma_{1,M}}{d\phi_P}
		+
		{\cal O}(\epsilon_2^2).
		\label{eq:sm-dsigma}
	\end{equation}
	The corresponding total-rate and second-harmonic moments are
	\begin{subequations}
		\begin{align}
			\Sigma_{i,M}
			&=
			\int d\phi_P\,
			\frac{d\sigma_{i,M}}{d\phi_P},
			\label{eq:sm-Sigma}
			\\
			W_{i,M}
			&=
			\int d\phi_P\,
			\cos2(\phi_P-\phi_\Delta)
			\frac{d\sigma_{i,M}}{d\phi_P},
			\qquad i=0,1 .
			\label{eq:sm-W}
		\end{align}
	\end{subequations}
	The normalized elliptic correlation is therefore
	\begin{equation}
		A_2^{(M)}
		=
		\frac{
			W_{0,M}+\epsilon_2 W_{1,M}
		}{
			\Sigma_{0,M}+\epsilon_2\Sigma_{1,M}
		}
		+
		{\cal O}(\epsilon_2^2).
		\label{eq:sm-A2-ratio}
	\end{equation}
	Expanding the ratio gives
	\begin{subequations}
		\begin{align}
			A_2^{(M)}
			&=
			v_{2,0}^{(M)}
			+
			\epsilon_2 S_M
			+
			{\cal O}(\epsilon_2^2),
			\label{eq:sm-A2-linear}
			\\
			v_{2,0}^{(M)}
			&=
			\frac{W_{0,M}}{\Sigma_{0,M}},
			\label{eq:sm-v20}
			\\
			S_M
			&=
			\frac{W_{1,M}}{\Sigma_{0,M}}
			-
			v_{2,0}^{(M)}
			\frac{\Sigma_{1,M}}{\Sigma_{0,M}} .
			\label{eq:sm-SM}
		\end{align}
	\end{subequations}
	The second term in Eq.~\eqref{eq:sm-SM} follows from the normalization of the angular correlation.  The same elliptic deformation modifies both the weighted numerator and the total diffractive rate.  Therefore a zero of $S_M$ is not equivalent to a zero of the diffractive cross section.
	
	The finite-rate null discussed in the main text is defined by
	\begin{equation}
		S_M(q_T,R_\gamma)=0,
		\qquad
		\Sigma_{0,M}(q_T,R_\gamma)\neq0.
		\label{eq:sm-null}
	\end{equation}
	It is a zero of the normalized linear response to the elliptic component, not a zero of the target dipole amplitude or of the diffractive rate.

	\section{Linearity check}
	\label{sec:sm-linearity}
	
\begin{figure}[t]
		\centering
		\includegraphics[width=0.65\linewidth]{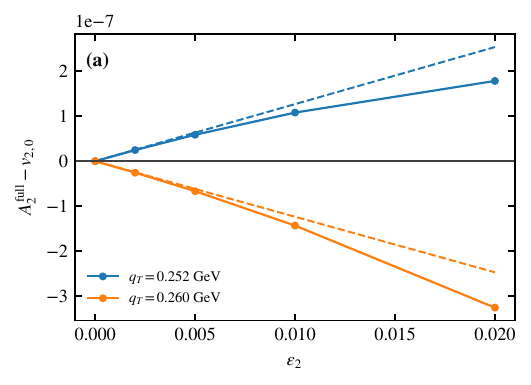}
		\caption{
			Linearity check of the OAM-resolved response.
			The full result $A_2^{\rm full}-v_{2,0}$ is shown as a function of $\epsilon_2$ for two representative values of $q_T$ near the $M=3$ finite-rate null.
			The dashed lines denote the first-order prediction $\epsilon_2 S_M$.
		}
		\label{fig:sm-linearity}
	\end{figure}

	The linear-response extraction is checked by computing the full elliptically deformed result for several values of $\epsilon_2$ and comparing it with the first-order prediction
	\begin{equation}
		A_2^{\rm full}-v_{2,0}^{(M)}
		=
		\epsilon_2 S_M
		+
		{\cal O}(\epsilon_2^2).
		\label{eq:sm-linearity}
	\end{equation}
	Figure~\ref{fig:sm-linearity} shows this comparison for two representative values of $q_T$ near the $M=3$ finite-rate null.  The agreement verifies that the zero shown in the main text is not a finite-difference artifact.  Higher-order terms in $\epsilon_2$ or kinematic smearing from a finite wave packet may shift the detailed zero position, but they do not alter the definition of the leading normalized response in Eq.~\eqref{eq:sm-SM}.
	
	\section{Stability of the finite-rate response zero}
	\label{sec:sm-robustness}
	
	The stability of the finite-rate response zero was tested by varying the parameters entering the benchmark dipole input and the transverse OAM projection one at a time.  The zero position is not expected to be universal, since it is determined by the overlap between the OAM--Bessel kernel and the transverse target profile.  The relevant stability criterion is instead whether the sign change of $S_3(q_T)$ persists while the baseline rate remains finite.  As shown in Table~\ref{tab:sm-robustness}, the response zero remains at a sizable fraction of the maximum baseline rate under representative variations of $B_0$, $\gamma$, $B_p$, and $R_\gamma$.  The largest displacement occurs for $B_p\to1.2B_p$, where the zero moves outside the narrow benchmark window and is recovered in a wider scan at lower $q_T$.

	\begin{table}[htbp]
		\centering
		\caption{
			Stability of the $M=3$ finite-rate response zero under representative one-at-a-time variations of the benchmark input and of the OAM wave-packet width.
			The quantity $R_\Sigma$ is defined as
			$R_\Sigma=\Sigma_{0,3}(q_T^\ast)/\max_{q_T}\Sigma_{0,3}(q_T)$
			within the corresponding scan range.
			All entries pass the V3 linear-response consistency check.
			The dagger marks the case for which a wider scan in $q_T$ is used because the zero is shifted outside the narrow benchmark window.
		}
		\begin{ruledtabular}
			\begin{tabular}{lccc}
				Variation
				&
				$q_T^\ast~({\rm GeV})$
				&
				$\Sigma_{0,3}(q_T^\ast)$
				&
				$R_\Sigma$
				\\
				\hline
				Baseline
				&
				0.256
				&
				0.062
				&
				0.839
				\\
				$B_0\to0.8B_0$
				&
				0.254
				&
				0.059
				&
				0.836
				\\
				$B_0\to1.2B_0$
				&
				0.259
				&
				0.064
				&
				0.841
				\\
				$\gamma\to0.9\gamma$
				&
				0.265
				&
				0.069
				&
				0.850
				\\
				$\gamma\to1.1\gamma$
				&
				0.248
				&
				0.056
				&
				0.830
				\\
				$B_p\to0.8B_p$
				&
				0.336
				&
				0.117
				&
				0.913
				\\
				$B_p\to1.2B_p^{\dagger}$
				&
				0.156
				&
				0.008
				&
				0.502
				\\
				$R_\gamma\to0.8R_\gamma$
				&
				0.248
				&
				0.050
				&
				0.828
				\\
				$R_\gamma\to1.2R_\gamma$
				&
				0.260
				&
				0.068
				&
				0.842
				\\
			\end{tabular}
		\end{ruledtabular}
		\label{tab:sm-robustness}
	\end{table}

\end{document}